\documentclass[aps,pra,superscriptaddress,amsmath,amssymb,preprintnumbers,floatfix,showpacs,12pt]{revtex4}
\usepackage{amssymb}
\usepackage{epsfig}

\begin{document}
\title{ Entanglement dynamics of electron-electron scattering in low-dimensional  semiconductor systems}
\author{F.~Buscemi }
\email{buscemi.fabrizio@unimore.it}
\author{P.~Bordone}
\affiliation{CNR-INFM, S3 Research Center, Via Campi 213/A, I-Modena 41100, Italy}
 \affiliation{ Dipartimento di Fisica, Universit\`{a} di Modena e Reggio Emilia, Modena, Italy}
\author{A.~Bertoni }
\affiliation{CNR-INFM, S3  Research Center, Via Campi 213/A, I-Modena 41100, Italy}
\begin{abstract}
We perform  the quantitative
evaluation of the entanglement dynamics in scattering events between
two insistinguishable electrons interacting via Coulomb potential in 1D and 2D
semiconductor nanostructures. We apply a criterion based on the
von Neumann entropy and the Schmidt decomposition of the global
state vector suitable for systems of identical  particles.
From the time-dependent numerical solution of the two-particle wavefunction
of the scattering  carriers we compute their entanglement evolution
for different spin configurations: two electrons with the same
spin, with different spin, singlet, and triplet spin state.
The procedure allows to evaluate the mechanisms that govern  entanglement
creation and their connection with the characteristic physical
parameters and initial conditions of the system. The cases in which
the evolution of entanglement is similar to the one obtained for
distinguishable particles are discussed. 
\end{abstract}
\maketitle

\section{Introduction} 
 
Quantum entanglement has undoubtly represented
one of the most peculiar aspects  of quantum mechanics as  it
can be viewed as the furthest departure of  quantum world from the
classical one \cite{Schr,Giuli,Peres}. Furthermore,  in the last years,  entanglement has been recognized as the resource  for  quantum information processing  \cite{Ved,Vog}. In this
context  a growing interest in continuous-variables entanglement  is arising for various  physical phenomena, such as binary
collisions between unbound or trapped
 particles, photoionization of atoms in both weak and strong fields,
and photodissociation of molecules, where the entangled subsystems are spatially separated \cite{Tal,Mark,Feredov,Eise}. For binary collisions  a particular attention has been devoted
 to the dynamical creation of entanglement between the two particles, stemming from the possibility
of using such a system to investigate the foundations of quantum mechanics itself: it has been shown that an almost
 maximal violation of Bell's  inequality   can occur \cite{Tzu}. The quest for quantum-computing  capable devices has also produced great interest in entanglement
formation in solid state systems where decoherence effects, mainly
due to carrier-carrier scattering, play a dominant role
\cite{Bertoni,Bordone,Gun}. In particular these effects have been
studied in the context of electron transport in semiconductors,
where the time evolution of the entanglement of spatial degrees of
freedom has been calculated numerically for  scattering events
between two distinguishable charged particles by means of the von
Neumann entropy  of    the  density matrix   reduced on the
spatial degrees of freedom of  one of the particles
\cite{Bertoni,Bordone}. It should be noticed   that   the
distinguishability of the two particles greatly simplifies the
system modelling and that it is  representative of only few of the
collision events taking place in semiconductors, as an
electron-hole scattering. Indeed  the notion of entanglement for
quantum systems composed of distinguishable particles has  been
widely investigated  from  the  theoretical and experimental point
of view \cite{Feredov,tso}, while   relevant questions arise for
systems of identical particles (both bosons and
fermions)\cite{Zanardi,Sch,Fisher,Shi,Li}. Here,  in particular,
unlike the distinguishable particles case, the Hilbert space of
the physical states does not have a naturally selected structure
of
 tensor product, because of the symmetrization postulate. For example,  in the analysis of a scattering event between particles, this implies  that
  the system  wavefunction
can  never be      factorized in the product of two one-particle wavefunctions. This makes difficult to give
a straightforward  definition of entanglement because of the presence of the
correlations due  to the exchange symmetry.

Only recently the problem has  been deeply analyzed  for fermion
systems. Various approaches   have been proposed in literature  \cite{Zanardi,Sch,Fisher,You,Vacca,Zanardi2}, each of one having its own advantages
and drawbacks, and no consensus  has been reached on which one is the most suitable \cite{brand}.
Here we follow the one introduced by Schliemann \cite{Sch,schli2} and recently  reexamined and extended by other authors \cite{You,ghira,sand},
since it allows to use a density matrix in the spatial coordinates and to compute the von Neumann entropy by means
of a generalization of the numerical procedure already developed for the case of distinguishable particles. Such an  approach 
 is based on  an  analogous of the Schmidt
decomposition for state vectors of two fermionic particles: through an
unitary transformation the antisymmetric wavefunction is
expressed into a basis of Slater determinants with a minimum
number of non-vanishing terms. This number, known as \emph{Slater rank}, is a criterion to identify whether a system is entangled or not, which involves  the evaluation
of  the von Neumann
entropy of the one-particle reduced density matrix. In this notation the von Neumann entropy
becomes not  only a measure of the  uncertainty due  to the impossibility  of attributing  a  definite state
to each particle of the couple  but also of the amount
of uncertainty deriving from the indistinguishability of particles. However it is well known that the quantum
correlations related to exchange simmetry  cannot be used to violate Bell's inequality
and are not a resource for quantum information processing \cite{ghira}: they  cannot therefore  be considered as a manifestation of \emph{genuine} quantum entanglement.

The basic significance of a quantitative evaluation of particle-particle entanglement comes also from the link between quantum correlations and decoherence \cite{Giuli}.
The latter is, in fact, a manifestation of the former when one drops the exact quantum description of the whole system, considering a part of it as the \emph{environment},
thus tracing out its quantum coordinates. In our case one can think of one of the particles as representing the environment.
As a consequence the amount of entanglement determined by the proposed simulative approach is a direct measure of the decoherence undergone by a carrier as a consequence of a carrier-carrier scattering.

In the present  paper  we aim at calculating entanglement of both
spatial and  spin degrees of freedom by studying, via a time-dependent numerical analysis, 1D and 2D models of a scattering
event between two electrons  in semiconductors. The  two particles,
explicitly  considered as indistinguishable, interact through a
Coulomb potential. This approach allows  us to overcome the hypothesis of
distinguishabilty of the particles used in  previous works \cite{Bertoni,Bordone}. In the present  case quantum entanglement must be analyzed
more carefully in order to single-out the effects of the additional correlation induced by indistinguishability. We also  evaluate the role of the spin degrees of freedom. We believe
that  our   results  can be a valuable  guideline for
 the  use of collision  events of   identical particles in
the quantum information processing based on  solid state systems and for the estimation of the decoherence
rate induced by carrier-carrier scatterings.

The paper is organized as follows. In Sec.~\ref{rte} we describe  the  method
used  to quantify entanglement in  two-fermion systems. In Sec.~\ref{rte1d}
we investigate the entanglement dynamics   for the case of a couple of electrons travelling in two parallel, single-mode, quantum wires (1D-system)
 and  in Sec.~\ref{rte2d} the same analysis is performed  for  a scattering event between   a free propagating and a bound electron  in a single-mode two-dimensional electron gas (2D-system).
Finally in Sec.~\ref{rtec}
 we comment the obtained results  and draw conclusions.

\section{Entanglement evaluation for two-fermion systems } \label{rte}
Here we shall briefly  describe the notion of entanglement  between   two fermions and the approach used for its evaluation, following  the theory
formulated by Schliemann \cite{Sch,schli2}. Any pure two-fermion  state can be written  as:
\begin{equation}
|\Psi_{F}\rangle=\sum_{i,j}^{2N}w_{ij}a^{\dag}_{i}a^{\dag}_{j}| 0\rangle \quad \quad \{a_{i},a^{\dag}_{j}\}=\delta_{ij}
\end{equation}
where $w_{ij}$ are the elements of a complex and antisymmetric ($2N\times2N$) matrix $W$,   and  $2N$ is the number of modes for each single particle.  $ a^{\dag}_{i}( a_{i})$  is  the creation(annihilation)
operator of a fermion in the $i$-th mode. The normalization condition
 is $\textrm{Tr}[W^{\dag}W]=1$.
Now we need to introduce the   fermionic analog of the Schmidt decomposition \cite{Met} indicating  that exists a unitary transformation
 $U$ such that
\begin{equation}\label{dsa}
W=UZU^{T}
\end{equation}
where
\begin{equation}
Z=\textrm{diag}
[Z_{1}\ldots Z_{i}\ldots Z_{N}]
\end{equation} 
is a block diagonal matrix with  bidimensional blocks  $Z_{i}$   of the type:
\begin{displaymath}
\mathrm{Z}_{i}=\left ( \begin{array}{cc}
0 & z_{i} \\
-z_{i}  &  0 \\
\end{array}\right ).
\end{displaymath}
We note that   $ z_{i}$ may also be zero. This decomposition is unique and we can  write
\begin{equation}\label{elio}
|\Psi_{F}\rangle=
\sum_{i}^{N}z_{i}(a'^{\dag}_{2i-1}a'^{\dag}_{2i}-a'^{\dag}_{2i}a'^{\dag}_{2i-1})| 
0\rangle
\end{equation}
where the fermionic state is given  in terms of new creation operators defined from  $a^{\dag}_{i}=\sum_{j}u^{\ast}_{ij}a'^{\dag}_{j}$. In other words 
one can think of the unitary transformation $U$    as acting   on the 
fermionic operator $a^{\dag}_{i}$.
The number of  $ z_{i}$ different from zero is the
Slater rank  and is closely related to the entanglement: the
state $|\Psi_{F}\rangle$ is considered  non entangled if  its
Slater rank  is equal to 1, while
it must be considered  a genuinly entangled state if  its Slater number is greater than
1. In fact it can be easily seen  from Eq.~(\ref{elio}) that  only in
the first  case
 $|\Psi_{F}\rangle$ can be obtained by antisymmetrization of the product of the single-particle states.

 It has been shown  recently that the  von Neumman entropy, commonly  used in the context of non-identical particles,
is still a good quantum-correlation measure for identical particles \cite{ghira,ghira2}. From the two-particle density matrix $\rho_{F}=|\Psi_{F}\rangle\langle\Psi_{F}|$, we can get
the one-particle reduced density matrix (normalized to unity)\cite{You}
\begin{equation}\label{fiu}
\rho_{\mu \nu}^{f}=\frac{\textrm{Tr}[\rho_{F}a_{\nu}^{\dag}a_{\mu}]}{
\textrm{Tr}[\rho_{F}{\sum_\mu}a_{\mu}^{\dag}a_{\mu}]}=(W^{\dag}W)_{\nu,\mu}.
\end{equation}
In terms of  the coefficients  $z_{i}$  the normalization condition  becomes     $ \sum_{i=1}^{N}|z_{i}|^{2}=1/2$, while
the  von Neumman  entropy $\varepsilon$ can be computed from:
\begin{equation}\label{ricci}
\varepsilon=-\textrm{Tr}[\rho^{f}\ln{\rho^{f}}]=\ln{2}-2\sum_{i=1}^{N}|z_{i}|^{2}\ln{2|z_{i}|^{2}}
\end{equation}
where, as stated before,  $|z_{i}|^{2}$ are the eingenvalues of the matrix
$\rho^{f}$. For identical particles the  von Neumann entropy reaches
its minimum value $\varepsilon=\ln{2}$ in the case of
non entangled states, whose   Slater rank  is  1, while,
for distinguishable particles, the minimum value of the entropy, still corresponding to non entangled states, is zero.
 This behavior can be explained if we consider  the real meaning of the von Neumman entropy, that
  must be interpreted as a measure of the lack of knowledge about the state of a quantum system. The value $\ln{2}$ for
 non entangled two-fermion states   measures the uncertainty arising
from the indistinguishability of the particles;  in this case
the correlations are only related to exchange simmetry  and
cannot be used as a resource for  quantum information. On the other side when the von Neumann
entropy is greater than $\ln{2}$
  there is an additional ignorance about the state of one of the two particles, stemming from  genuine entanglement.

\section{1D-system}\label{rte1d}
\subsection{The Theoretical  Model}
The physical system  consists of two   electrons   running
in opposite directions along two parallel silicon quantum wires.
The dynamics of the two  particles  is 1D and is coupled  through  
 Coulomb potential, so that the Hamiltonian of the system
takes the form
\begin{equation}\label{rest}
H(r_{a},r_{b})= - \frac{\hbar^{2}}{2m}\left(\frac{\partial^{2}}{\partial r_{a}^{2}}+\frac{\partial^{2}}{\partial r_{b}^{2}}\right)+V(r_{a},r_{b})
\end{equation}
where $V(r_{a},r_{b})$  can be expressed as
 \begin{equation}
V(r_{a},r_{b})=\frac{e^{2}}{\epsilon \sqrt{\left(x_{a}-x_{b}\right)^{2}+d^{2}}}
\end{equation}
where $e$ is the electron charge, $\epsilon$ is the silicon
dielectric constant, $ x_{a}$ and   $ x_{b}$ the  electron
coordinates  along the wires and  $d$   the distance in  the
perpendicular  direction $y$  between  the two wires. We  use the approximation
 of single parabolic band and do not consider spin-orbit interaction.

\begin{figure}[h]
  \begin{center}
    \includegraphics[width=0.3\textwidth]{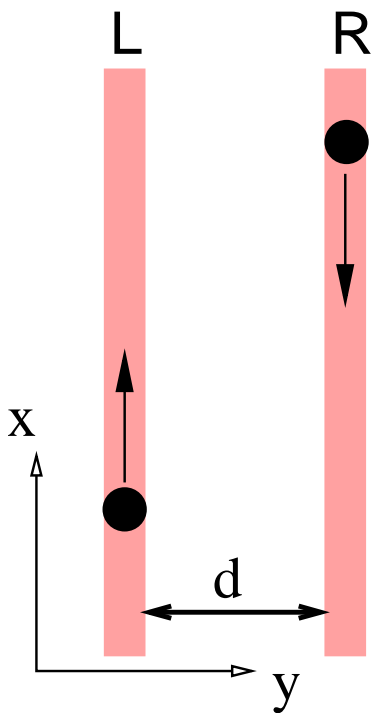}
   \caption{\label{fig:conW} (Color online) The 1D physical system consisting of  two parallel quantum wires,
   where two electrons run in opposite directions; $d$ is the
   distance between the  wires.}
 \end{center}
\end{figure}

The  two quantum wires have been considered as 1D-systems. Each of
the two particles
 has another  degree of freedom, besides the spin and space one, indicating the quantum wire in which it is
 moving, i.e. we suppose that each   carrier is  confined  along  the $y$
direction  in a  wire  (see Fig. \ref{fig:conW}).
 Moreover each of the two carriers is represented at initial time $t_{0}$ by a minimum uncertainty wave packet, so that
  its   space wavefunction    takes  the  form
\begin{equation} \label{rosi}
\psi(x,t_{0})= \frac{1}{(\sqrt{2 \pi}\sigma)^{1/2}}\exp{\left( -\frac{(x-x_{0})^{2}}{4(\sigma)^{2}}+i k\cdot x\right)}
\end{equation}
where  $\sigma$ is the mean dispertion in position, $k=\sqrt{2 m E_{k}}/\hbar $ with  $m$  the effective mass of
the carrier and $E_{k}$ its kinetic energy.

Taking  into account both spin and  ``wire'' degrees of freedom, we can consider different
initial    state configurations. The first quantum state considered is the one having
 two particles with   the same spin (spin up):
\begin{equation}\label{res}
|\Psi\rangle=\frac{1}{\sqrt{2}}\bigg(|\psi \,\phi\rangle |\textsf{L}\, \textsf{R} \rangle   - |\phi \,\psi\rangle |\textsf{R}\, \textsf{L}\rangle\bigg)|\!\!\uparrow \uparrow \rangle
\end{equation}
where  both  the  wavefunctions corresponding to the states $|\psi\rangle$ and $|\phi\rangle$ are of the type defined in Eq.~(\ref{rosi}), and   the   variances of the wavepackets  and the
 distance between their  centers are  such that the Coulomb energy of the system is negligible at initial time.
In Eq.~(\ref{res}),  the ket  $|\textsf{L}\rangle (|\textsf{R} \rangle)$  indicates that the particle is localized  in left (right) wire,
and   $|\!\! \uparrow \rangle$ indicates   spin up state.

 In order to compute the entanglement as given by Eq.~(\ref{ricci})  we need to evaluate numerically the quantum state of  Eq.~(\ref{res}).
To this aim the explicit form of the matrix $W$ has to be obtained. This is done  in the Appendix \ref{appe}, where we show
that  $W$ for the same-spin configuration   takes the form
\begin{equation}\label{ficci}
\mathrm{W_{\Psi}}=\frac{1}{2\sqrt{2}}\left ( \begin{array}{cccc}
W_{A} & - W_{S} & 0 &  0\\
W_{S}  & - W_{A} &  0 &  0\\
0  & 0  & 0  &  0 \\
0  & 0  & 0  & 0 \\
\end{array}\right )
\end{equation}
where $W_{A} $ and  $W_{S} $ are the antisymmetric and symmetric
matrices defined in the Appendix \ref{appe}. 

 The second quantum state considered is the one
 of two electrons having different spins. It is not
factorizable in a part containing only space and wire
variables and in a part containing only spin variables,  and reads
\begin{equation} \label{drog}
   |\Upsilon \rangle=\frac{1}{\sqrt{2}}    \bigg(|\psi\, \phi\rangle   |\textsf{L} \, \textsf{R} \rangle |\!\! \uparrow  \downarrow \rangle  - |\phi\, \psi\rangle   | \textsf{R} \,\textsf{L} \rangle
|\!\! \downarrow \uparrow \rangle\bigg).
\end{equation}
It should be noticed  that  for such a  state  it is possible  to make precise
claims about  the spin   of a particle whose position-wire
state is known.
From  the calculations presented  in the Appendix \ref{appe} we get
\begin{equation}\label{gicci}
\mathrm{W_{\Upsilon}}=\frac{1}{4\sqrt{2}}\left ( \begin{array}{cccc}
W_{A} & - W_{S} &  - W_{S} &  W_{A}\\
W_{S}  & - W_{A} &  - W_{A} &  W_{S}\\
W_{S}  & - W_{A} &  - W_{A} &  W_{S}\\
W_{A} & - W_{S} &  - W_{S} &  W_{A}\\
\end{array}\right ).
\end{equation}

The last  two  states  considered still describe   two electrons with different spins, but they 
 can be factorized in a position-wire  coordinate term and in a spin  term. We can identify  the singlet spin state
\begin{eqnarray}\label{quest}
|\Phi\rangle=\frac{1}{2}\bigg(|\psi \,\phi\rangle |\textsf{L}\, \textsf{R} \rangle   + |\phi \,\psi\rangle | \textsf{R}\, \textsf{L}\rangle\bigg)
\bigg(| \!\!\uparrow \downarrow  \rangle -|   \!\!\downarrow \uparrow \rangle\bigg)
\end{eqnarray}
with
\begin{equation}\label{hicci}
\mathrm{W_{\Phi}}=\frac{1}{4}\left ( \begin{array}{cccc}
0 & 0 &  - W_{S} &  W_{A}\\
0  & 0 &  - W_{A} &  W_{S}\\
W_{S}  & - W_{A} & 0 & 0 \\
W_{A} & - W_{S} & 0 &  0 \\
\end{array}\right ),
\end{equation}
and the triplet  spin state
\begin{equation}\label{quest1}
|\Xi\rangle=\frac{1}{2}\bigg(|\psi \,\phi\rangle |\textsf{L}\, \textsf{R} \rangle   - |\phi \,\psi\rangle | \textsf{R}\, \textsf{L}\rangle\bigg)
\bigg(|\!\!\uparrow \downarrow  \rangle +|\!\!\downarrow \uparrow \rangle\bigg)
\end{equation}
with
\begin{equation}
\mathrm{W_{\Xi}}=\frac{1}{4}\left ( \begin{array}{cccc}
W_{A} & - W_{S} &  0 & 0\\
W_{S}  & - W_{A} & 0 &  0\\
0  & 0 &  - W_{A} &  W_{S}\\
0 &0 &  - W_{S} &  W_{A}\\
\end{array}\right ).
\end{equation}
The triplet and singlet  states  have  relevant properties that have  been  taken into account
in evaluating the von Neumann entropy. In fact, as shown in the Appendix \ref{appe},   the  von Neumann entropy
 of  $|\Xi\rangle$ and $|\Phi\rangle$ can  be  obtained by adding $\ln{2}$ to  the one of the same-spin state $|\Psi\rangle$.
This property, due to the fact that such states  are factorizable, remains true also during
the  time evolution since    the  Hamiltonian
 does not include  spin terms. This implies that at any  time $t$  the triplet and singlet   can  always be    factorized in
a position-wire  term and  a spin term. Therefore we can attribute  the  offset  of the entanglement value to the lack of knowledge
about the spin state, in the sense that one  cannot
make precise claims about the spin state of the particle in a given position-wire  state
$|\psi  \textsf{L}\rangle $ or $|\phi  \textsf{R}\rangle $  \cite{ghira}.
\subsection{Numerical results}
In our approach we solve numerically the time-dependent Schr\"{o}ndiger equation for the two-particle
 spatial  wavefunction   by means of a Crank-Nicholson  finite difference scheme. Once  the spatial  wavefunctions
are obtained at each time step,  we calculate the matrices    $W$'s related to  the different initial spin state configurations   and,  from these, 
we obtain the one-particle reduced density matrixes $\rho$'s. This allows us to evaluate the time evolution of the entanglement
between the spatial   and  spin degrees of freedom  of the two electrons in terms of the  von Neumann entropy according  to the expression
 given in Eq.~(\ref{ricci}).  To this purpose we need to diagonalize the one-particle reduced density matrix at each time step
and this turns out to be the most  demanding  part of our  algorithm from the point of view of the numerical calculation.

\begin{figure}[h]
  \begin{center}
    \includegraphics[width=0.8\textwidth]{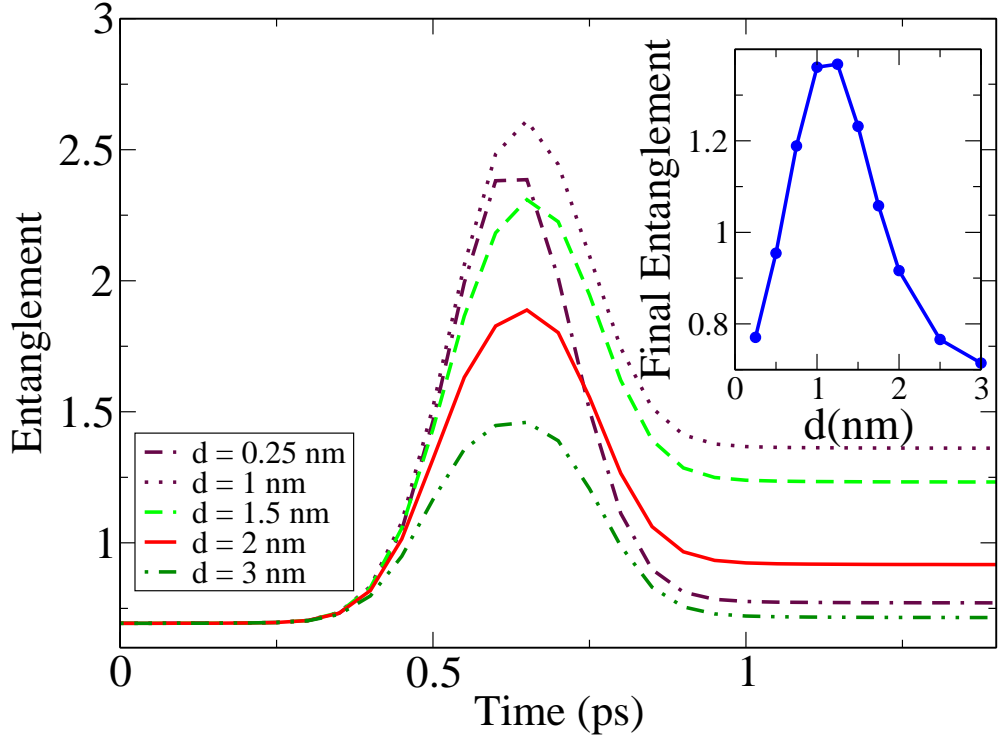}
   \caption{\label{fig:con2}  (Color online)  Entanglement as a function   of time for the  state $|\Upsilon\rangle$ (see Eq.~(\ref{drog})) for 
different values of the distance $d$ between the two wires. The two electrons have the same initial energy  $E_{k}$ = 50 meV and
are described by two wavepackets with an 
initial mean dispersion $\sigma$ = 20 nm. The inset shows the stationary values
of the entanglement as a function of $d$. }
 \end{center}
\end{figure}

\begin{figure}[h]
  \begin{center}
    \includegraphics[width=0.8\textwidth]{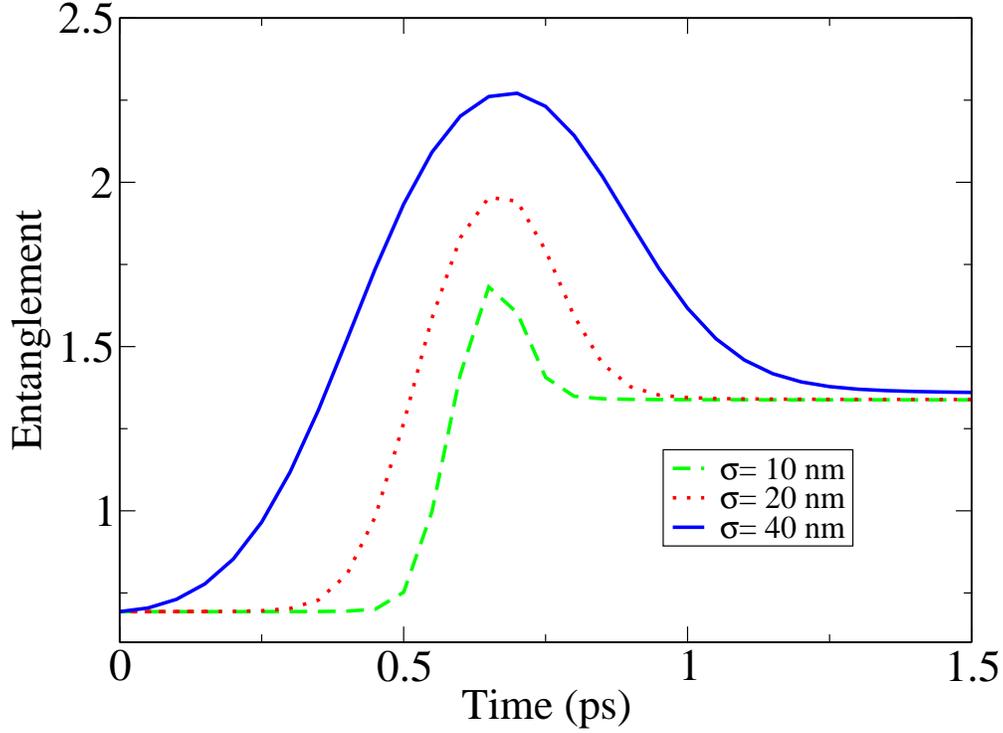}
   \caption{\label{fig:con3} (Color online) Entanglement as a function   of time for the same-spin state $|\Psi\rangle$   at different values of the 
 initial mean dispersion $\sigma$   of the  spatial  wavefunctions. The  initial kinetic energy  $E_{k}$ is  50 meV for both    electrons. Note that the final  stationary value is pratically independent
upon  $\sigma$.}
 \end{center}
\end{figure}

Figs.~\ref{fig:con1} and \ref{fig:con2} show  that,  at initial time, when the two particles are far enough  and Coulomb interaction is  negligible, the von Neumann entropy
for the states $|\Psi\rangle$ and $|\Upsilon\rangle$  is $\ln{2}$   as expected \cite{ghira,ghira2} .
\begin{figure}[h]
  \begin{center}
    \includegraphics[width=0.8\textwidth]{buscemi_fig2.eps}
   \caption{\label{fig:con1} (Color online) Entanglement as a function   of time for the  state $|\Psi\rangle$  desribing two electrons 
with the same spin for  different values of the distance $d$ between the two wires. The two electrons have the same initial energy  $E_{k}$ = 50 meV and
are described by two wavepackets with an
initial mean dispersion $\sigma$ = 20 nm. The inset shows the stationary values
of the entanglement as a function of $d$. }
 \end{center}
\end{figure}
As indicated by  the theoretical criterion described
in Sec.~\ref{rte}  this value can be related to the 
unavoidable correlations due to exchange symmetry and does not imply 
  genuine quantum entanglement. In fact     $|\Psi\rangle$ and $|\Upsilon\rangle$ are
initially  non entangled. The entanglement increases while the
two electrons are approaching each other and finally reaches a
stationary value once electrons get apart. The final entanglement, which  is a function of  the distance $d$ between the two wires,
 is reported in the insets of Figs.~\ref{fig:con1} and \ref{fig:con2}. Following this approach  it is
possible to obtain a quantitative evaluation of the  time
of  entanglement formation in a scattering event: for the studied case such a time is found to be around  1
picosecond. Obviously the dynamics of entanglement formation
depends on the physical parameters of the system, as the  initial kinetic energy of the  electrons   and the variance $\sigma$ of their initial wavepacket.
This is illustrated in Fig.~\ref{fig:con3} where $\varepsilon(t)$  is reported   for different values  of the variance $\sigma$.
 Increasing  the mean dispersion in position leads to an increase in the time at which  the stationary value of
entanglement  is reached, but   the  value itself
 is pratically not affected.

We note that the dependence on $d$ of the final values   of the
entanglement  is significantly different for the case of the state
$|\Psi\rangle$  with respect to the case of the state
 $|\Upsilon\rangle$. While for $|\Psi\rangle$ this value decreases monotonically as  the distance between
 the two wires is increased, for
$|\Upsilon\rangle$ it shows  a  well defined maximum (around $d=1$ nm  in Fig.~\ref{fig:con2} inset). This means that,   in the case of
electrons having   the same spin,  the scattering creates strong
quantum  correlations between the particles    which are  weakly
affected by  the wire  distance. On the other hand for electrons
with different spins we find  a strong  dependence on $d$ of
the stationary values of the entanglement. This behavior is in
good qualitative agreement with the one obtained  in previous works 
in the case of distinguishable particles.

\begin{figure}[h]
  \begin{center}
    \includegraphics[width=0.8\textwidth]{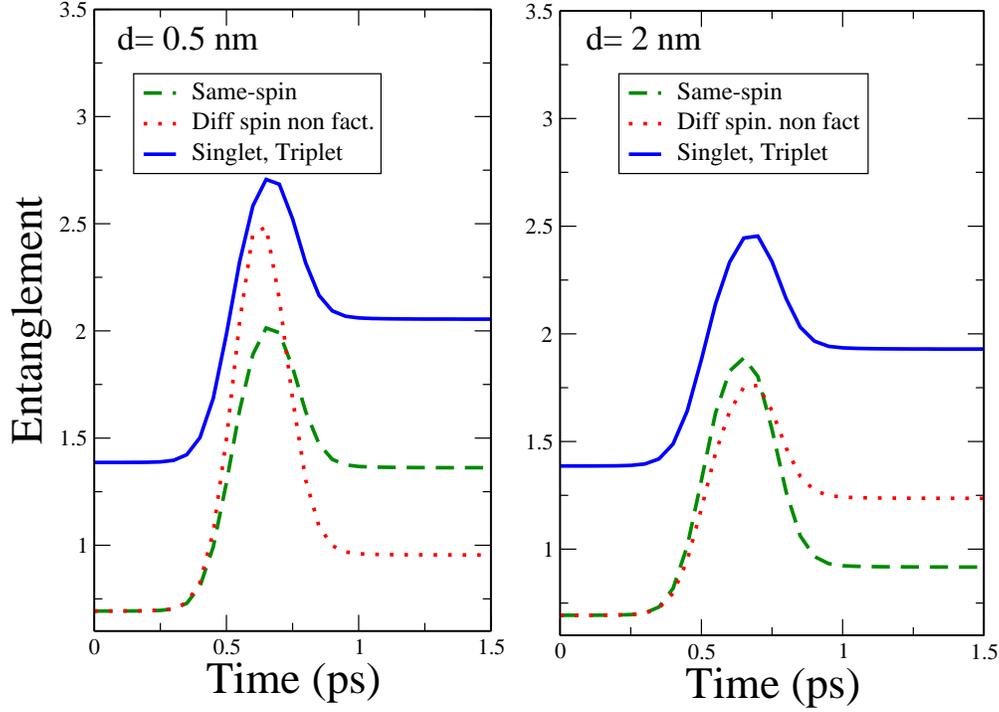}
   \caption{\label{fig:con4} (Color online) Entanglement as a function   of time: two electrons with the same spin $|\Psi \rangle$(dashed lines), with different spin $|\Upsilon \rangle$(dotted line),
singlet $|\Phi \rangle$ and triplet $|\Xi \rangle$  (solid line).
 The two wavepackets describing the electrons have both  an initial energy $E_{k}$ = 50 meV and initial mean dispersion $\sigma$ = 20 nm. Two different values of the distance
 $d$ between the two wires are considered:  d = 0.5 nm  in the left panel and d = 2 nm in the right panel.}
 \end{center}
\end{figure}

In Fig.~\ref{fig:con4} we report the  time evolution of the entanglement for two values of the distance between the wires and for  all the considered
  spin configurations. Intially  the   von Neumann entropy of the singlet $|\Phi\rangle $ and triplet  $|\Xi\rangle$ states  has the value  $2\ln{2}$. In  agreement
 with the theoretical predictions   its   time evolution   results to be the same obtained  for  the same-spin state $|\Phi\rangle$,
 just  shifted by $\ln{2}$. Thus  at the initial time
the singlet and triplet states  have to  be considered as 
entangled  according to the notion of entanglement for fermionic
particles used in this work. The appearance of the offset above mentioned  can be explained considering the mathematical form
of  $|\Phi \rangle$ and   $|\Xi \rangle$ as given in Eqs.~(\ref{quest}) and (\ref{quest1}). Their
Slater rank is greater than  1 since they cannot be put in
terms of a single Slater determinant: in fact they are not
obtained by antisymmetrization of the product of single-particle
states. This also  confirms that the  von Neumann entropy, as a
measure of entanglement, corresponds to the   Slater rank
criterion \cite{ghira,You}.

\section{2D-System }\label{rte2d}
\subsection{The theoretical model }
We study in this Section the entanglement formation for the case of an electron  propagating
in a 2D system and interacting with another electron bound to a specific site by a harmonic potential.  The latter   can be considered as a simple model of
 a  shallow impurity in a semiconductor. Here  the physical parameters
of GaAs have been used. The dynamics of the particle bound
in the  harmonic potential is coupled to the incoming particle through
a 
 Coulomb interaction. The Hamiltonian reads
\begin{equation}\label{rest2}
H(\textbf{r}_{a},\textbf{r}_{b})= - \frac{\hbar^{2}}{2m}\left(\frac{\partial^{2}}{\partial \textbf{r}_{a}^{2}}+\frac{\partial^{2}}{\partial \textbf{r}_{b}^{2}}\right)+\frac{e^{2}}{\epsilon |\textbf{r}_{a}-\textbf{r}_{b}|}+\frac{1}{2}m\omega^{2}(\textbf{r}_{a}-\textbf{r}_{0})+\frac{1}{2}m\omega^{2}(\textbf{r}_{b}-\textbf{r}_{0})
\end{equation}
where $\epsilon$ and $m$ are  the GaAs dielectric constant and  effective mass, respectively, and $\textbf{r}_{0}$ is the  center  of the harmonic potential,
with  energy-level spacing  $\hbar\omega$.

At the initial time $t_{0}$ one of the two particles (incoming
electron) is represented by a minimum uncertainty wavepacket $\psi(\textbf{r},t_{0})$ centered in $\textbf{r}_{1}$, with $\textbf{k}$ vector pointing towards $\textbf{r}_{0}$, 
 while   the other (bound electron) is in the ground state of the 2D harmonic oscillator
\begin{equation} \label{bisc1}
\phi(\textbf{r},t_{0})= \left(\frac{m\omega}{\pi \hbar}\right)^{1/2}\exp{\left( -\frac{m\omega(\textbf{r}-\textbf{r}_{0})^{2}}{2\hbar}\right)}.
\end{equation}
The distance $|\textbf{r}_{1}-\textbf{r}_{0}|$ is such that   at  the initial time $t_{0}$  the Coulomb energy
  is negligible. Moreover  we have taken $\sigma=\sqrt{\hbar / 2m\omega}$ : this
 allowed us to evaluate   the time evolution of the two-particle   wavefunction  in  the center-of-mass coordinate system.

Let us now consider  the state $|\Psi\rangle$ describing the system with  the two particles having  the same spin. At the initial time
\begin{equation} \label{mefi}
|\Psi\rangle=\frac{1}{\sqrt{2}}\bigg(|\psi \,\phi\rangle    - |\phi \,\psi\rangle \bigg)|\!\!\uparrow \uparrow \rangle.
\end{equation}
When we discretize  each space dimension with a grid of N points we obtain
\begin{equation}
|\Psi\rangle=\frac{1}{\sqrt{2}}\sum_{i,j}^{N^{2}}\bigg(\psi( \textbf{r}_{i})\,\phi(\textbf{r}_{j})    - \phi(\textbf{r}_{i}) \,\psi(\textbf{r}_{j}) \bigg)
| \textbf{r}_{i}\textbf{r}_{j}\rangle |\!\!\uparrow \uparrow \rangle
\end{equation}
and applying the same formalism used for the  1D  system of  Section \ref{rte1d}  we get for the  $2N^{2} \times 2N^{2}$ matrix $W$ associated to the same-spin state
\begin{equation}
\mathrm{W_{\Psi}}=\frac{1}{\sqrt{2}}\left( \begin{array}{cc}
W_{A} & 0\\
0  &  0 \\
\end{array}\right),
\end{equation}
where $W_{A}$ is the antisymmetric $N^{2} \times N^{2}$  matrix
whose elements  read
\begin{equation}
w_{ij}=\psi( \textbf{r}_{i})\,\phi(\textbf{r}_{j})    - \phi(\textbf{r}_{i}) \,\psi(\textbf{r}_{j}),
\end{equation}
 obtained by the antisymmetrization of the product of the
single-particle wavefunctions $\psi$ and $\phi$. The other 
spin configuration we study  in this Section  is the triplet spin state
\begin{equation}
|\Xi\rangle=\frac{1}{2}\bigg(|\psi \,\phi\rangle    - |\phi \,\psi\rangle \bigg)\bigg(|\!\!\uparrow \downarrow\rangle+|\!\!\downarrow\uparrow \rangle\bigg)
\end{equation}
for which 
\begin{equation}
\mathrm{W_{\Xi}}=\frac{1}{2}\left( \begin{array}{cc}
W_{A} & 0\\
0  &  -W_{A} \\ 
\end{array}\right).
\end{equation}
Also in this case the von Neumann entropy of the triplet   $|\Xi\rangle$ is
obtained by the one of the same-spin state $|\Psi\rangle$ adding $\ln{2}$.
The singlet state  and  the state  with the electrons having different spins
and non factorizable into a spin term and a space term have not been  considered in the 2D case, since they need a much larger computational effort that is beyond the scope 
of the present work.
\subsection{Numerical results }
The  time evolution of the 2D two-particle
wavefunction  $|\Psi\rangle$  has been performed in the center-of-mass coordinate system, by means of a finite-difference
solution of the corresponding time-dependent Schr\"odinger
equation \footnote{In order to scale  the time  of computation and
to have a  memory space big enough  to allocate the matrices involved in the calculations, we have developed
a  parallel algorithm  for  matrix diagonalization.} .

\begin{figure}[h]
  \begin{center}
    \includegraphics[width=0.8\textwidth]{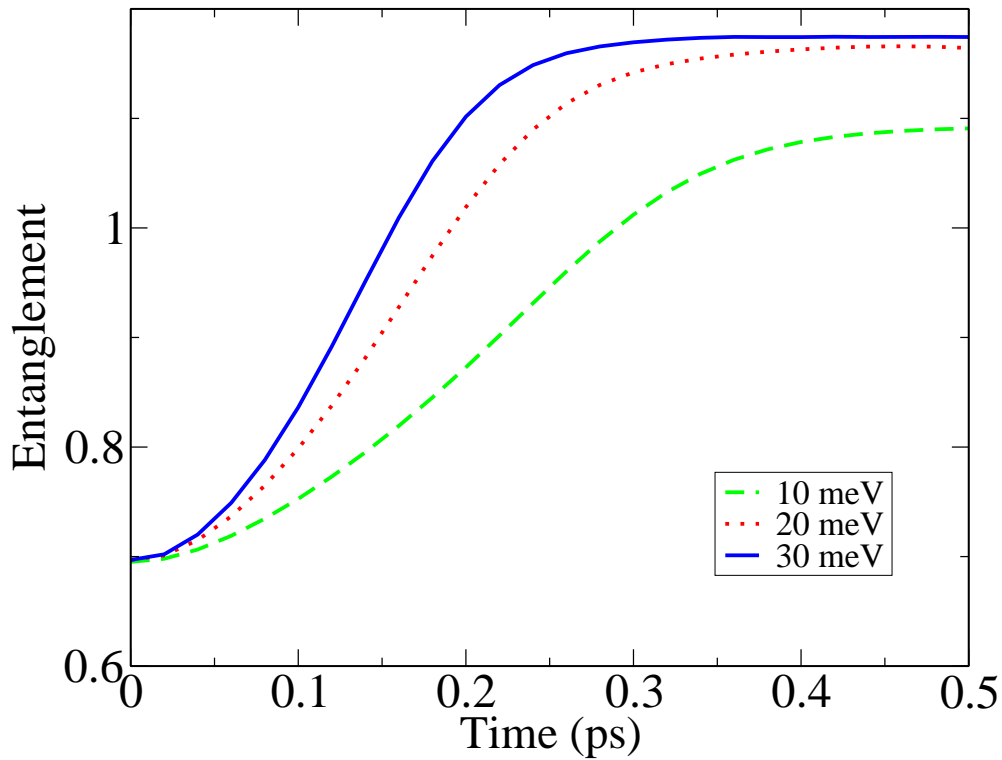}
   \caption{\label{fig:con5} (Color online) The time evolution of entanglement for $|\Psi\rangle$ (see Eq.~(\ref{mefi}))  in the 2D system at three different values
of the  initial energy of the incoming electron, namely $E_{k}$ = 10 meV (dashed line), 20 meV (dotted line), 30 meV (solid line). The harmonic oscillator energy is $\hbar\omega=2$ meV.}
 \end{center}
\end{figure}

At initial time the  centers of the  wavepackets are distant enough so
that the Coulomb  energy is negligible:
as a consequence,  when the system is described by  the   state  $|\Psi\rangle$  (the two electrons have the same spin),  no correlation
is  present  apart from the one  due  to the exchange symmetry. In fact we see from Fig.~\ref{fig:con5} that  for   any value of the energy of the incoming electron
 the  initial  von
Neumman entropy gets  its minimum  value $\ln{2}$. As the
particles get closer their quantum correlation  builds  up and
entanglement reaches a stationary value once particles are far enough. Here the final amount of entanglement depends
on the initial energy of the carrier:  it is  higher  for higher energies.
The time evolution of $\varepsilon(t)$  for the triplet  state
can be obtained  simply shifting  by $\ln{2}$ the curves reported in Fig.~\ref{fig:con5}.

\begin{figure}[h]
  \begin{center}
    \includegraphics[width=0.8\textwidth]{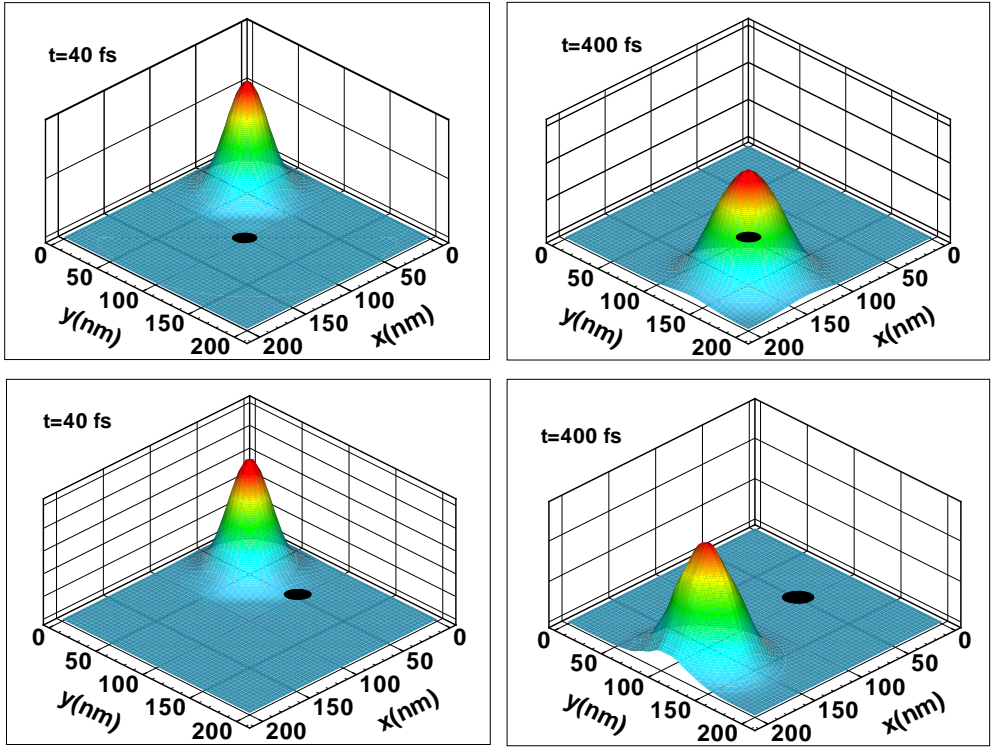}
   \caption{\label{fig:con6} (Color online) The square modulus of the two-particle  antisymmetrized  wavefunction
\mbox{$\psi( \textbf{r})\,\phi(\textbf{R})    - \phi(\textbf{R}) \,\psi(\textbf{r})$}  for $|\Psi\rangle$ at two different times:
 $t=40$ fs  and  $t=400$ fs. In the two upper  graphs the position of one of two particles is fixed
to \textbf{R}=(100 nm, 100 nm), while in the two lower graphs \textbf{R}=(50 nm, 150 nm).}
 \end{center}
\end{figure}

Fig.~\ref{fig:con6} is intended to give a better insight into the
role played by the correlations between the space degrees of
freedom in the entanglement creation, for the  $|\Psi\rangle$
state case. To this purpose  we have calculated the square modulus
of  the wavefunction, at two different times, keeping the
position of  one of the two electrons fixed (indicated by
the black spot in the figures). The first time (left column)
corresponds to a condition of minimum entanglement, the second (right column) to a condition in which
entanglement has already reached its stationary values. We note
that when entanglement is at its lowest level  the result is independent on
the choice of the fixed position  while this is not
true  once entanglement has been created by the Coulomb
interaction.

\section{Conclusions}\label{rtec}
In this paper  we have studied the  dynamics of entanglement 
formation in systems composed by two indistinguishable particles.
To this aim we applied    a   theoretical method
\cite{Sch,schli2}, which permits the quantitative  evaluation of
entanglement for a two-fermion system. The method  is based on a
 fermionic analog of the Schmidt decomposition of the  state vector and agrees with
von Neumman entropy of the  single-particle reduced density
matrix, which is   a  good correlation measure and allows one to 
determine  whether the lack of knowledge is due to a genuine 
entanglement or only to  particles indistinguishabilty. It should
be noticed that  the correlations due to exchange symmetry  cannot
be used to implement  quantum information processing or violate
Bell's inequality \cite{ghira,ghira2}.

The  time-dependent numerical implementation of the
 theoretical method   we developed  allows one  to investigate the
decoherence effects that are consequence of carrier scattering,
and that can be quantified in terms of entanglement between the scattered particles. This approach
makes feasible a direct quantification of the time of the entanglement formation. The numerical results have
been obtained  using  the     effective mass of  electrons in
semiconductors (Si for the 1D model and GaAs for the 2D model) but
 they   can be considered as representative of a  general behavior.

Our simulations show that  the final value of entanglement between
two electrons undergoing a  scattering event  depends  only upon
some specific physical parameters,  as the distance between the
 quantum wires (in the 1D case), or  the  initial  kinetic energy of the
particles, while it is essentially immune to the modifications of other
parameters, like the variance $\sigma$ of the initial
wavepackets. We showed how the  entanglement  dynamics depends
on  the spin components  of the state  vector describing  the
system   at the initial time,  even if the Hamiltonian does not
contain  spin terms. This suggests the existence of a deep
relation between spin states and space entanglement, which implies the
possibility of  detecting quantum correlations  in spin
measurements  even for the case of  wavepackets extending in two distant, not overlapping,  regions of space. Moreover, in the 1D  system, the numerical results  for singlet and triplet spin states
shows,  at the initial time,  quantum correlations
related  not only  to exchange simmetry but also to  genuine
entanglement. They  can be attributed to the  lack of knowledge  about
the spin state of a  particle in a specific wire, in agreement  with theoretical
predictions.

 The results
obtained  in the case of two electrons having different spins are 
in good qualitative agreement with the case of  distinguishable 
particles \cite{Bertoni,Bordone} apart from the offset term  
 deriving from   exchange simmetry.
 This  also confirms that  the
theoretical framework   given in   Refs.~\cite{Sch,You,ghira,sand}
 is a suitable   criterion   for the  evaluation of the entanglement in two-fermion systems and is equivalent, in the case of distinguishable particles
to the commonly used criterion based on  the Schmidt number or on
the von Neumman entropy.

\begin{acknowledgments}
We are pleased to thank Carlo Jacoboni for fruitful discussions.
We  acknowledge support from the U.S Office of Naval Research (contract No. N00014-03-1-0289/ N00014-98-1-0777) and INFM-CNR
Progetto Supercalcolo 2006 CINECA.
\end{acknowledgments}

\appendix*
\section{Determination of W  Matrices  } \label {appe}
Here  we illustrate the  procedure  we apply  to   the  different  initial  states in order to evaluate numerically the
von Neumann entropy in 1D physical systems with each electron localized in the left ($|\textsf{L}\rangle$) or right($|\textsf{R}\rangle$) wire, as explained in Sec.\ref{rte1d}.
Let us begin with  the state $|\Psi\rangle$ defined in Eq. (\ref{res})
where   both  particles have spin  up but  different position-wire  wavefunctions  described by $|\psi\rangle|\textsf{L}\rangle$
or $|\phi\rangle|\textsf{R}\rangle$. We   discretize the $x$ coordinate of the  electrons along the two wires with  a N points grid:
the dimension of  each single-particle Hilbert space is $2N$,   taking into account the spin  degrees of freedom. In this way   we can consider  a fermionic position
operator $a^{\dag}(x_{l})$  that  creates  a particle in the point $x_{l}$:
$a^{\dag}(x_{l})| 0 \rangle = | x_{l} \rangle $.

We can then express the state $|\Psi\rangle$ as
\begin{eqnarray} \label{reggia}
|\Psi\rangle &=&\frac{1}{\sqrt{2}}\sum_{l,m}^{N} \bigg(\psi(x_{l})\phi(x_{m})  |\textsf{L} \,\textsf{R} \rangle   - \phi(x_{l})\psi(x_{m})
|\textsf{R}\, \textsf{L}\rangle\bigg)\nonumber \\
&\times& |x_{l}x_{m}\rangle|\!\!\uparrow \uparrow  \rangle .
\end{eqnarray}
We observe that in Eq.~(\ref{reggia})  the state $|\Psi\rangle$ is not antisymmetric with respect to the exchange of  the $l$ and $m$  indeces,
and, in this form,  it cannot be used   to evaluate the entanglement according to  the theory developed in Sec.~\ref{rte}. This  difficulty  can be overcome
by   applying  a unitary transformation to  the  wire degrees of freedom:
\begin{equation}\label{mattafix}
 \begin{array}{ccc}
|V_{1}\rangle &=&1/\sqrt{2} \bigg(|\textsf{L}\rangle +|\textsf{R}\rangle\bigg)\\
|V_{2}\rangle&=&1/\sqrt{2} \bigg(|\textsf{L}\rangle -|\textsf{R}\rangle\bigg).\\
\end{array}
\end{equation}

Inserting  the closure relation  in Eq.~(\ref{reggia}), we get
\begin{eqnarray} \label{bott}
|\Psi\rangle &=&\frac{1}{\sqrt{2}}\sum_{l,m}^{N}\sum_{s,t=1}^{2}\bigg(\psi(x_{l})\phi(x_{m}) \gamma_{\textsf{L}}(s) \gamma_{\textsf{R}}(t)   -
 \phi(x_{l})\psi(x_{m})\gamma_{\textsf{R}}(s) \gamma_{\textsf{L}}(t)\bigg) \nonumber \\
&\times&|x_{l}x_{m}\rangle|V_{s}V_{t}\rangle|\!\!\uparrow \uparrow  \rangle
\end{eqnarray}
where we introduced  $\gamma_{\textsf{L}}(1)=\gamma_{\textsf{L}}(2)=
\gamma_{\textsf{R}}(1)=\frac{1}{\sqrt{2}} $ and  $\gamma_{\textsf{R}}(2)=
-\frac{1}{\sqrt{2}}$.

Now we can express the state $|\Psi \rangle $ in Eq.~(\ref{bott}) in terms of
fermionic operators  $f_{i}^{\dag}$ defined as:
 \begin{displaymath}
f_{i}^{\dag}| 0 \rangle = \left \{ \begin{array}{cccc}
|x_{i}\rangle |V_{1}\rangle|\!\! \uparrow \rangle & & &\textrm{for} \quad 1\le i \le N  \\
|x_{i-N}\rangle |V_{2}\rangle|\!\! \uparrow\rangle  & & &\textrm{for} \quad N+1\le i \le 2N \\
|x_{i-2N}\rangle|V_{1}\rangle|\!\! \downarrow \rangle  & & &\textrm{for} \quad 2N+1\le i \le 3N  \\
|x_{i-3N}\rangle|V_{2}\rangle|\!\! \downarrow\rangle  & & &\textrm{for} \quad 3N+1\le i \le 4N  \\
\end{array}\right .
\end{displaymath}
and we get
\begin{equation}\label{lezio}
|\Psi\rangle= \sum _{i ,j }^{4N} w_{ij} f_{i}^{\dag}f_{j}^{\dag} |0\rangle
\end{equation}
where the  $4N\times 4N$ matrix  of the  coefficients $  w_{ij}$  takes the form of Eq.~(\ref{ficci}),
where   $W_{A}$  is an  antisymmetric   $N \times N$
matrix whose
 elements  read
\begin{equation}\label{piss}
w_{A _{lm}}=\psi(x_{l})\phi(x_{m})-\phi(x_{l})\psi(x_{m})
\end{equation}
and $W_{S}$  is  its symmetric conterpart. They have been
 obtained by antisymmetrizing or symmetrizing the product of the single-particle spatial  wavefunctions $\phi$ and $\psi$
calculated in the grid points $l$ and $m$.

Now the  whole matrix $W_{\Psi}$ is antisymmetric
and  the state $|\Psi\rangle$ in
Eq.~(\ref{lezio}) is expressed in a form that allows us to  apply the  Schliemann   theory. The normalization condition
$\textrm{Tr}[W_{\Psi}^{\dag}W_{\Psi}]=1$ is satisfied. Moreover during the time evolution $W_{\Psi}$ mantains this form since the Hamiltonian does not contain spin terms and  the  coefficients
 corresponding to the  modes where particles have spin down will  always be
 zero.

For the  initial state $|\Upsilon\rangle$ of Eq.~(\ref{drog}),  with  two electrons having  different spins and not
factorizable in a part containing only space and ``wire'' variables and in a part containing only spin variables,
  we again discetize the $x$  coordinate  into a $N$  points grid
and consider the change of  basis given in Eq.~(\ref{mattafix}).
 In this case we   have to apply a unitary transformation also for the spin variables:
 from  $|\!\! \uparrow \rangle$ and $ |\!\! \downarrow \rangle$ to $|S_{1}\rangle$ and $|S_{2}\rangle$
\begin{equation}
\begin{array}{ccc}
|S_{1}\rangle&=&1/\sqrt{2} (|\!\! \uparrow\rangle +|\!\! \downarrow\rangle)\\
|S_{2}\rangle&=&1/\sqrt{2} (|\!\! \uparrow\rangle -|\!\! \downarrow\rangle).\\
\end{array}
\end{equation}
Now we can express $|\Upsilon\rangle$ as
\begin{eqnarray}
 |\Upsilon\rangle=\frac{1}{\sqrt{2}} \sum_{l,m}^{N}\sum_{s,t,u,v=1}^{2}\bigg(\psi(x_{l})\phi(x_{m}) \gamma_{\textsf{L}}(s) \gamma_{\textsf{R}}(t)\beta_{\uparrow}(u)  \beta_{\downarrow}(v)   - \nonumber \\
 \phi(x_{l})\psi(x_{m}) \gamma_{\textsf{R}}(s) \gamma_{\textsf{L}}(t)  \beta_{\downarrow}(u)\beta_{\uparrow}(v)   \bigg) |x_{l}x_{m}\rangle|V_{s}V_{t}\rangle  |S_{u}S_{v}\rangle
\end{eqnarray}
where $\beta_{\uparrow}(1)=\beta_{\uparrow}(2)=\beta_{\downarrow}(1)= \frac{1}{\sqrt{2}}$ and $\beta_{\downarrow}(2)= -\frac{1}{\sqrt{2}}$.

If we define the fermionic  creation  operators  $g_{i}^{\dag}$ as:
 \begin{displaymath}
g_{i}^{\dag} |0 \rangle = \left \{ \begin{array}{cccc}
|x_{i}\rangle |V_{1}\rangle |S_{1}\rangle  & & &\textrm{for} \quad 1\le i\le N  \\
|x_{i-N}\rangle |V_{2}\rangle |S_{1}\rangle  & & &\textrm{for} \quad N+1\le i\le 2N  \\
|x_{i-2N}\rangle |V_{1}\rangle |S_{2}\rangle  & & &\textrm{for} \quad 2N+1\le i\le 3N  \\
|x_{i-3N}\rangle |V_{2}\rangle |S_{2}\rangle  & & &\textrm{for} \quad 3N+1\le i\le 4N  \\
\end{array}\right .
\end{displaymath}
the state  $|\Upsilon\rangle$ in Eq.~(\ref{drog}) can be expressed as
\begin{equation}\label{cuci}
|\Upsilon\rangle= \sum _{i,j}^{4N} w_{ij}g_{i}^{\dag}g_{j}^{\dag} |0\rangle.
\end{equation}
The  $4N\times 4N$  matrix $W_{\Upsilon}$ reads as in Eq.~(\ref{gicci}).

Finally  we are able to apply    to  the singlet spin state of Eq. (\ref{quest}) and to triplet spin state  of Eq. (\ref{quest1})
 the same procedure used above for  $|\Upsilon\rangle$, and  we find that
$|\Phi\rangle$ is given in an expression  similar to  Eq.~(\ref{cuci}),
for which   the matrix $W$  takes the  form of Eq.~(\ref{hicci}), while for the state $|\Xi\rangle$
$W$ takes the form of   Eq.~(\ref{quest1}). For the sake of brevity we do not report explicitely those calculations.

Now we shall show that the von Neumann  entropy $\varepsilon_{\Xi}$ of the triplet  $|\Xi\rangle$ and the singlet $|\Phi\rangle$  states (both representing the electrons with opposite spin)
 can  easily be  related to the entropy $\varepsilon_{\Psi}$
of the same-spin state $|\Psi\rangle$. Firstly we  note that $|\Xi\rangle$ and $|\Phi\rangle$
have the same entropy: this depends on the  fact that the one-particle reduced density matrix $W^{\dag}W$
results to be  the same  in the two cases. Secondly we observe that the
 matrix ${W_{\Xi}}$ can be viewed  as
a block diagonal matrix with 2  blocks of dimension $2N$.
Therefore   for the eingenvalues  $|z_{i}|^{2}$ of  $W_{\Xi}^{\dag}W_{\Xi}$   holds the relation 
\begin{equation}\label{voxa}
|z_{i}^{\Xi}|^{2}=|z_{i+2N}^{\Xi}|^{2}.
\end{equation}
In this case   the  $\varepsilon_{\Xi}$ can be written as:
\begin{equation}\label{tras}
\varepsilon_{\Xi}=\ln{2}-2\sum_{i=1}^{2N}|z_{i}^{\Xi}|^{2}\ln{2|z_{i}^{\Xi}|^{2}}.
\end{equation}
The matrix $W_{\psi}$ has the same upper left block as   $W_{\Xi}$ and is zero elsewhere. This implies that the   first $2N$ eingenvalues of the matrix  $W_{\psi}^{\dag}W_{\psi}$
   are equal (apart from the constant factor $\frac{1}{2}$ coming from the normalization condition)  to  the   first $2N$
eingenvalues of the matrix  $W_{\Xi}^{\dag}W_{\Xi}$. Using  (\ref{voxa}) and (\ref{tras})  we can relate
the von Neumann entropy for  $|\Xi\rangle$  to the one for $|\Psi_{}\rangle$
\begin{equation}\label{trasf}
\varepsilon_{\Xi}=\ln{2}-4\sum_{i=1}^{N}\frac{1}{2}|z_{i}^{\Psi}|^{2}\ln{2\frac{1}{2}|z_{i}^{\Psi}|^{2}} = \varepsilon_{\Psi}+ \ln{2}
\end{equation}
where we have taken into account the normalization condition
$\sum_{i=1}^{N}|z_{i}^{\Psi}|^{2}=\frac{1}{2}$,  recalling
$z_{i}^{\Psi}=0$ for $i>2N$. Eq. (\ref{trasf}) implies that the von
Neumann entropy of the  triplet spin state can be simply  obtained
from the one of same-spin state by adding $\ln{2}$.


\begin{thebibliography}{99}

\bibitem{Giuli}
D.~Giulini et al., \emph{Decohence and the Appearance of a Classical World in Quantum theory}
(Springer, Berlin, 1996.)




\bibitem{Peres}
A.~Peres,  \emph{Quantum Theory: Concepts and Methods}
(Kluwer Academy Publishers, The Netherlands, 1995.)




\bibitem{Schr}
E.~Schrodinger,
  Naturwissenschaften  \textbf{23}, 807 (1935).



\bibitem{Ved}
V.~Vedral, M.B.~Plenio, M.A. Rippin and P.L Knight,
 Phys. Rev. Lett. \textbf{78}, 2275 (1997).
\bibitem{Vog}
E.~Shchukin   and W.~Vogel,
 Phys. Rev. Lett. \textbf{95}, 230502 (2005).

\bibitem{Tal}
A.~Tal  and G.~Kurizki,
 Phys. Rev. Lett.  \textbf{94}, 160503, (2005)


\bibitem{Mark}
H.~Mack  and M.~Freyberger,
 Phys. Rev. A \textbf{66}, 042113 (2002).


\bibitem{Feredov}
M.V.~Fedorov \emph{et  al.},
Phys. Rev. A  \textbf{69}, 052117 (2004).


\bibitem{Eise}
J.~Eisert and M.B.~Plenio,
 Int. J. Quant. Inf \textbf{1}, 479 (2003).


\bibitem{Tzu}
Tzu-Chieh Wei  \emph{et  al.},
  Phys. Rev. A  \textbf{67}, 022110 (2003).



\bibitem{Bertoni}
A.~Bertoni,
 J.Comp.Elec. \textbf{2}, 291 (2003).


\bibitem{Bordone}
P.~Bordone and A.~Bertoni,
 J.Comp.Elec  \textbf{3}, 407 (2004).

\bibitem{Gun}
D.~Gunlycke, J.H.~Jefferson, T.~Rejec, A.~Ramsak, D.G.~Pettifor, G.A.D.~Briggs,
cond-matter/0509010 .

\bibitem{tso}
H.~Mack and M.~Freyberger,
 Phys. Rev. A  \textbf{66}, 042113 (2002).

\bibitem{Shi}
Yu~Shi,
 Phys. Rev. A, \textbf{67}, 024301 (2003).

\bibitem{Li}
Y.S.~Li, B.~Zeng, X.S.~Liu and G.L.~Long,
 Phys. Rev. A, \textbf{64}, 054302  (2001).


\bibitem{Zanardi}
  P.~Zanardi,
 Phys. Rev. A \textbf{65}, 042101(R)(2002).





\bibitem{Sch}
J.~Schliemann, J.I.~Cirac, M.~Kus, M.~Lewenstein and D.~Loss,
 Phys. Rev. A \textbf{64}, 022303 (2001).



\bibitem{Fisher}
J.R.~Gittings and A.J.~Fisher,
 Phys. Rev. A  \textbf{66}, 032305 (2002). 


\bibitem{You}
R.~Paskauskas and L.~You,
 Phys. Rev. A  \textbf{64}, 042310 (2001).


\bibitem{Vacca}
H.M.~Wiseman and J.A.~Vaccaro,
 Phys. Rev. Lett.  \textbf{91}, 097902  (2003).



\bibitem{Zanardi2}
P.~Zanardi and Wang X,
J. Phys. A: Math. Gen.  \textbf{35}, 7947 (2002).


\bibitem{brand}
Fernando G.S.L.~Brandao,
 Phys. Rev. A  \textbf{72}, 022310  (2005).








\bibitem{schli2}
J.~Schliemann, D.~Loss, and A.H.~MacDonald,
 Phys. Rev. B \textbf{63}, 085311 (2001).
 

\bibitem{ghira}
G.C.~Ghirardi and L.~Marinatto,
 Phys. Rev. A  \textbf{70}, 012109 (2004).

\bibitem{sand}
X.~Wang and B.~Sanders,
  J. Phys. A: Math. Gen. \textbf{38}, 67 (2005).



\bibitem{Schl3}
K.~Eckert, J.~Schliemann, D.~Bruss and  M.~Lewenstein,
Annals of Physics \textbf{299}, 88 (2002).

\bibitem{Met}
M.L.~Metha, {\em Elements of Matrix Theory\/} (Hindustian Publishing Corporation, Delhi,1977.)











\bibitem{ghira2}
G.~Ghirardi, L.~Marinatto and T.~Webber,
 J. Stat. Phys \textbf{108}, 49 (2002).






\end{thebibliography}
\end{document}